\documentclass{aa501}     
\usepackage{graphics}
\def\cc{\,{\rm cm^{-3}}}
\def\cm2{\,{\rm cm^{-2}}}
\def\kms{\,{\rm {km\,s^{-1}}}}
\def\kkms{\,{\rm {K\,km\,s^{-1}}}}
\def\h2{\,{\rm H_{2}}}
\def\13co{\,{\rm ^{13}CO}}
\def\co{\,{\rm ^{12}CO}}
\def\pci{\,{\rm ^{3}P_{1}-^{3}P_{0}\,[CI]}}
\def\pcii{\,{\rm ^{2}P_{3/2}-^{2}P_{1/2}\,[CII]}}
\def\mu{\,\mu m}
\def\aua{A\&A, }

\def\apj{ApJ, }

\def\apjl{ApJL, }

\def\mnras{MNRAS, }
\begin{document}
 
\title{Neutral atomic carbon in centers of galaxies}
 
   \subtitle{}
 
   \author{F.P. Israel
          \inst{1}
           and F. Baas
          \inst{1,2, \dag}
	   }
 
   \offprints{F.P. Israel}
 
  \institute{Sterrewacht Leiden, P.O. Box 9513, NL 2300 RA Leiden,
             The Netherlands
  \and       Joint Astronomy Centre, 660 N. A'ohoku Pl., Hilo,
             Hawaii, 96720, USA}

\date{\dag: Deceased April 4, 2001
\\
\\
Received ; Accepted}
 
\abstract{
We present measurements of the emission from the centers of fifteen 
spiral galaxies in the $\pci$ fine-structure transition at 492 GHz. 
Observed galaxy centers range from quiescent to starburst to active.
The intensities of neutral carbon, the $J$=2--1 transition of 
$\13co$ and the $J$=4--3 transition of $\co$ are compared in matched
beams. Most galaxy centers {\it emit more strongly in [CI] than in 
$\13co$}, completely unlike the situation pertaining to Galactic
molecular cloud regions. [CI] intensities are lower than, but nevertheless 
comparable to $J$=4--3 $\co$ intensities, again rather different from
Galactic sources. The ratio of [CI] to $\13co$ increases
with the central [CI] luminosity of a galaxy; it is lowest for
quiescent and mild starburst centers, and highest for strong starburst
centers and active nuclei. Comparison with radiative transfer model
calculations shows that most observed galaxy centers have {\it neutral 
carbon abundances close to, or exceeding, carbon monoxide abundances}, 
rather independent from the assumed model gas parameters. The same models
suggest that the emission from neutral carbon and carbon monoxide, 
if assumed to originate in the same volumes, {\it arises from a warm and
dense gas} rather than a hot and tenuous, or a cold and very dense gas.
The observed [CI] intensities together with literature [CII] line
and far-infrared continuum data likewise suggest that a significant
fraction of the emission originates in medium-density gas 
($n = 10^{3}-10^{4}\cc$), subjected to radiation fields of various 
strengths.
}

\maketitle
 
\section{Introduction}

%
\begin{figure*}[t]
\unitlength1cm
\begin{minipage}[]{16.4cm}
\resizebox{17cm}{!}{\rotatebox{270}{\includegraphics*{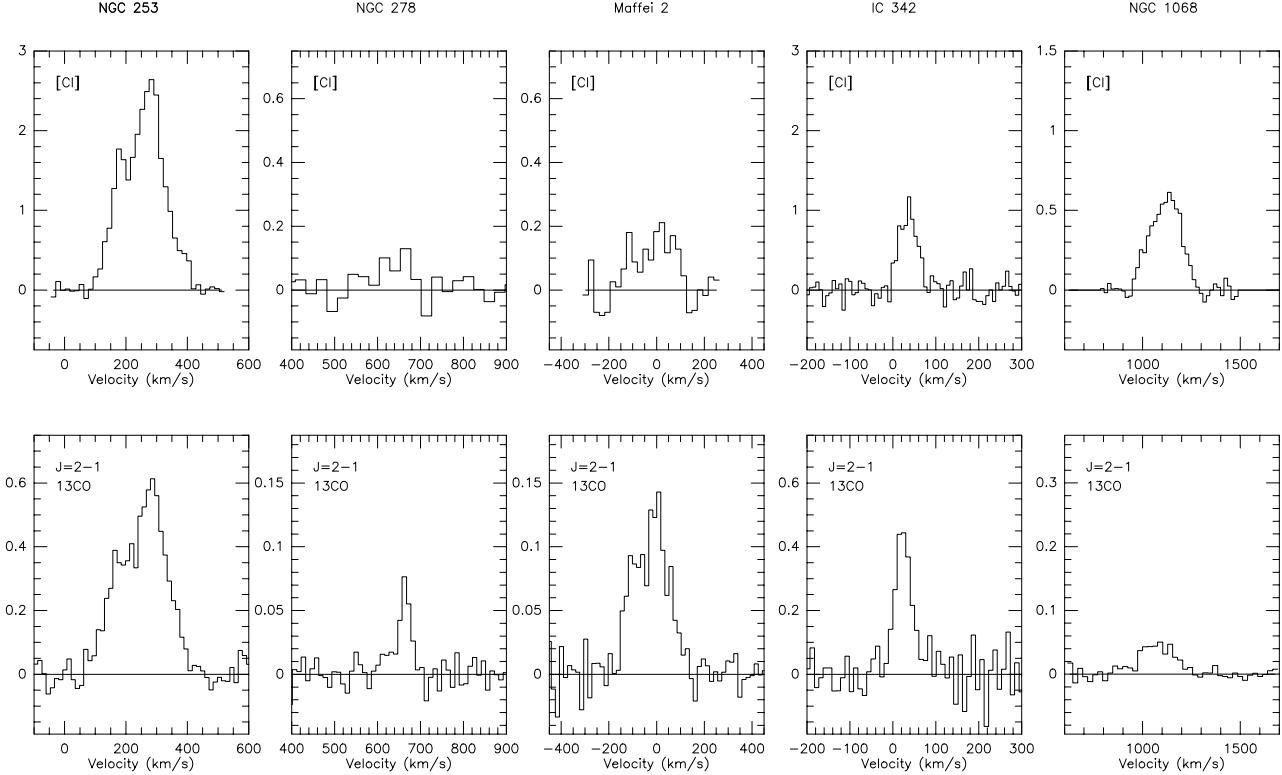}}}
\end{minipage}
\caption[]{[CI] and $J$=2--1 $\13co$ spectra observed towards 
sample galaxies. The vertical scale is $T_{\rm mb}$ in K; the 
horizontal scale is velocity $V_{\rm LSR}$ in $\kms$. For all
galaxies, the temperature range in [CI] is four times that in
 $\13co$. }
\end{figure*}

\bigskip

%
\begin{table*}
\caption[]{Line observations log}
\begin{center}
\begin{tabular}{lcccrcccccc}
\hline
\noalign{\smallskip}
Galaxy	 & \multicolumn{2}{c}{Position} & Adopted  & \multicolumn{3}{c}{[CI] } & \multicolumn{2}{c}{$J$=2--1$\13co$ } & \multicolumn{2}{c}{$J$=4--3 $\co$} \\
  & RA(1950) & Dec(1950)       & Distance & No. [CI] & Date  & $T_{\rm sys}$ & Date  & $T_{\rm sys}$ & Date & $T_{\rm sys}$\\
  & (h m s) & ($^{\circ}$ $'$ $''$) & (Mpc) & Points &	     & (K) 	     & 	     & (K) & & (K) \\
\noalign{\smallskip}
\hline
\noalign{\smallskip}
NGC~253  & 00:45:05.7 	& --25:33:38 &  2.5     & 20 & 12/93 & 3770	     & 12/93 & 1695 	     & 11/94 & 9800 \\
NGC~278  & 00:49:15.0	& +47:16:46  & 12       &  1 & 07/96 & 3650	     & 06/95 &  480 	     & 01/01 & 1325 \\
NGC~660  & 01:40:21.6	& +13:23:41  & 13       &  1 & 07/96 & 3065	     & 05/01 &	350 	     & 08/99 & 3870 \\
Maffei~2 & 02:38:08.5	& +59:23:24  &  2.7     &  1 & 12/93 & 4885	     & 01/96 &  550 	     & 07/96 & 3700 \\
NGC~1068 & 02:40:07.2	& --00:13:30 & 14.4     & 22 & 07/96 & 4000	     & 01/96 &  455	     & 07/96 & 3365 \\
IC~342   & 03:41:36.6	& +67:56:25  &  1.8     & 27 & 11/94 & 4485	     & 02/89 & 1440	     & 04/94 & 2170 \\
M~ 82	 & 09:51:43.9	& +69:55:01  &  3.25    &  6 & 12/93 & 7200	     & 04/93 &  335	     & 10/93 & 9085 \\
NGC~3079 & 09:58:35.4	& +55:55:11  & 18.0     &  7 & 03/94 & 6240	     & 06/95 &  310	     & 03/94 & 5510 \\
NGC~3628 & 11:17:41.6	& +13:51:40  &  6.7     &  8 & 11/94 & 3450	     & 06/95 &  325	     & 03/94 & 2414 \\
NGC~4826 & 12:54:17.4	& +21:57:06  &  5.1     &  2 & 03/97 & 3520	     & 12/93 &  535	     & 12/93 & 2045 \\
M~51	 & 13:27:45.3	& +47:27:25  &  9.6     &  4 & 11/94 & 6600	     & 06/95 &  370	     & 04/96 & 4065 \\
M~83	 & 13:34:11.3	& --29:36:39 &  3.5     & 14 & 12/93 & 4590	     & 06/95 &  430	     & 12/93 & 4360 \\
NGC~5713 & 14:37:37.6   & --00:04:34 & 21.0	&  1 & 02/99 & 8000	     & 12/00 &  515	     & ...   & ...  \\
NGC~6946 & 20:33:48.8	& +59:58:50  &  5.5     & 17 & 07/96 & 3970	     & 01/96 &  530	     & 07/96 & 2900 \\
NGC~7331 & 22:34:46.6	& +34:09:21  & 14.3     &  1 & 11/96 & 1925	     & 12/97 &  320	     & ...   & ...  \\
\noalign{\smallskip}
\hline
\end{tabular}
\end{center}
\end{table*}

\begin{figure*}[t]
\unitlength1cm
\begin{minipage}[]{16.4cm}
\resizebox{17.0cm}{!}{\rotatebox{270}{\includegraphics*{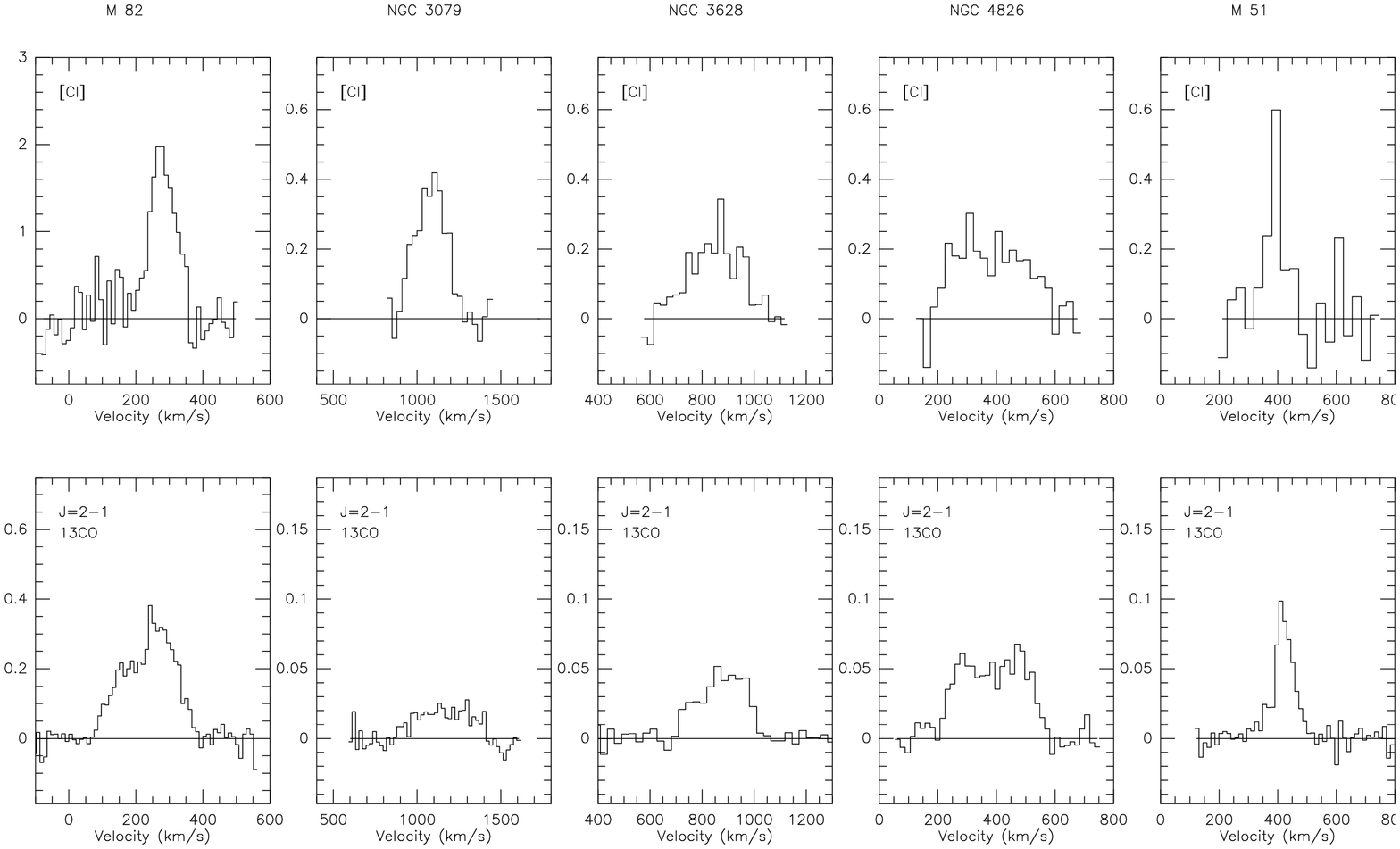}}}
\end{minipage}
\end{figure*}

\bigskip
%
\begin{table*}
\caption[]{[CI], $J$=2--1 $\13co$, $J$=4--3 $\co$ intensities}
\begin{center}
\begin{tabular}{lccccccc}
\hline
\noalign{\smallskip}
Galaxy & Offset & \multicolumn{5}{c}{Center Position}			 & Area-integrated \\
       &        & $T_{\rm mb}$([CI]) & $I$([CI]) & $I(\co 4-3)$ & $I$([CI]) & $I(\13co)$ & [CI] Luminosity \\
	 & & \multicolumn{3}{l}{(10$''$)}    & \multicolumn{2}{l}{(22$''$)} & \\
    & $''$ & (mK) & \multicolumn{4}{c}{($\kkms$)} 	      	 & ($\kkms$ kpc$^{2}$) \\
\noalign{\smallskip}
\hline
\noalign{\smallskip}
NGC~253	  & 0, 0     & 2615  & 486$\pm$60  & 1019$\pm$120 & 290$\pm$45  & 106$\pm$13    &   14$\pm$2.2     \\
NGC~278	  & 0, 0     &  100  &   7$\pm$3   &    9$\pm$2   &  (5$\pm$1)  &  2.6$\pm$0.4  &   (3.1$\pm$0.7)  \\
NGC~660	  & 0, 0     &  240  &  50$\pm$8   &   85$\pm$12  & (31$\pm$8)  &  7.8$\pm$1.1  &  (26$\pm$7)    \\
Maffei~2  & 0, 0     &  190  &  37$\pm$7   &  405$\pm$50  & (20$\pm$7)  &  22$\pm$4     &   (0.9$\pm$0.3) \\
NGC~1068  & 0, 0     &  560  & 109$\pm$12  &  153$\pm$19  &  49$\pm$9   &  11$\pm$2     &   53$\pm$8.5     \\
IC~342	  & 0, 0     & 1030  &  54$\pm$6   &  209$\pm$21  &  27$\pm$7   &  24$\pm$3     &    1.1$\pm$0.3   \\
M~82      & 0, 0     & 2130  & 224$\pm$35  &  591$\pm$95  & 180$\pm$30  &  60$\pm$9     &   39$\pm$6.9     \\
NGC~3079  & 0, 0     &  530  & 111$\pm$18  &  115$\pm$20  & (70$\pm$15) &  12$\pm$3     &  143$\pm$31     \\
NGC~3628  & -17, +5  &  265  &  84$\pm$11  &  110$\pm$15  &  38$\pm$8   &  10$\pm$2     &   28$\pm$5.7     \\
NGC~4826  & 0, 0     &  135  &  11$\pm$2   &  ...         &  17$\pm$4   &   7.8$\pm$1.6 &  ...     \\
          & -20, +5  &  270  &  86$\pm$10  &   73$\pm$9   & (47$\pm$10) &  15$\pm$2     &  (13$\pm$3)     \\
M~51      & 0, 0     &  565  &  28$\pm$5   &   24$\pm$4   & (13$\pm$4)  &   8.4$\pm$1.9 &  (8$\pm$2.7)     \\
	  & -12, -12 &  340  &  24$\pm$5   &  ...         & ...         &  ...          &   ...    \\
	  & -12, -24 &  755  &  55$\pm$9   &  ...         & ...         &  ...          &   ...    \\
	  & -24, -24 &  470  &  55$\pm$9   &  ...         & ...         &  ...          &   ...    \\
M~83	  & 0, 0     &  685  &  83$\pm$14  &  270$\pm$20  &  55$\pm$8   &  29$\pm$3     &    3.6$\pm$0.5   \\
NGC~5713  & 0, 0     & $<$90 &   2$\pm$0.4 &  ...         & ...         &   7.4$\pm$1.6 & 1.7--2.6 \\
NGC~6946  & 0, 0     &  465  &  85$\pm$9   &  216$\pm$20  &  44$\pm$8   &  22$\pm$3     &   13$\pm$2.4     \\
NGC~7331  & 0, 0     &   30  &   2$\pm$0.3 &	 ...      & ...         &   2.5$\pm$0.6 & 0.8--1.7 \\
\noalign{\smallskip}
\hline
\end{tabular}
\end{center}
Note: Offset position (--20,+5) of NGC~4826 is actual nucleus position.
\end{table*}

Carbon monoxide (CO), the most common molecule after $\h2$, is now
routinely detected in external galaxies. However, when exposed to
energetic radiation, CO is readily photodissociated turning atomic carbon 
into an important constituent of the interstellar medium. As the 
ionization potential of neutral carbon is quite close to the dissociation 
energy of CO, neutral carbon subsequently may be ionized rather easily. 
As a consequence, [CI] emission primarily arises from interface regions
between zones emitting in [CII] and CO respectively (see e.g. Israel et al.
1996; Bolatto et al. 2000). It requires column densities sufficiently
high for shielding against ionizing radiation, but not so high that
CO selfshielding allows most gas-phase carbon to be bound in molecules.
In principle, observations of emission from CO, C$^{\circ}$ and 
C$^{+}$ provide significant information on the physical condition of the 
cloud complexes from which the emission arises (Israel, Tilanus $\&$ Baas 
1998; Gerin $\&$ Phillips 2000; Israel $\&$ Baas 2001). Even though the
far-infrared continuum and the [CII] line are much more efficient coolants,
the various CO and [CI] lines are important coolants for relatively cool, 
dense molecular gas, contributing about equally to its cooling (Israel
et al. 1995; Gerin $\&$ Phillips 2000). In galaxies, however, studies of 
the dense interstellar medium are complicated by the effectively very large 
linear observing beams which incorporate whole ensembles of individual, 
mutually different clouds. The clumpy nature of the interstellar medium 
allows UV radiation to penetrate deeply into cloud complexes, so that the 
CO, [CI] and [CII] emitting volumes appear to coincide when observed with 
large beamsizes. The physics and structure of such photon-dominated 
regions (PDR's) has been reviewed most recently by Hollenbach $\&$ Tielens 
(1999), whereas their consequent observational parameters have been 
modelled by e.g. Kaufman et al. (1999).

[CII] emission has been observed towards numerous galaxies, both from 
airborne (the now defunct NASA Kuiper Airborne Observatory) and from spaceborne 
(the equally defunct Infrared Space Observatory) platforms. In contrast to these
[CII] observations, observations of [CI] emission can be performed on the ground, 
at least in the $\pci$ transition at 492 GHz. However, atmospheric transparency is 
poor at such high frequencies and weather conditions need to be unusually 
favourable for observations of the often weak extragalactic [CI] emission 
to succeed, even at the excellent high-altitude site
of telescopes as the JCMT and the CSO. Consequently, the number of published 
results is relatively limited. Beyond the Local Group, i.e. at distances larger
than 1 Mpc, [CI] has been
mapped in bright galaxies such as IC~342 (B\"uttgenbach et al. 1992), M~82 
(Schilke et al. 1993; White et al. 1994; Stutzki et al. 1997) and 
NGC~253 (Israel et al. 1995; Harrison et al. 1995), as well as M~83 
(Petitpas $\&$ Wilson, 1998; Israel $\&$ Baas 2001) and NGC 6946 
(Israel $\&$ Baas 2001). A survey of 13 galaxies, including limited radial 
mapping of NGC~891 and NGC~6946 was recently published by Gerin $\&$ 
Phillips (2000). In this paper, we present a similar [CI] survey of 15 
galaxies. We also obtained $J$=2--1 $\13co$ measurements for all galaxies,
and $J$=4--3 measurements for all but two. Taking overlap into account, 
this survey brings the total number of galaxies outside the Local Group, 
detected in [CI], to 26. 

\section{Observations}
All observations were carried out with the 15m James Clerk Maxwell
Telescope (JCMT) on Mauna Kea (Hawaii) \footnote{The James Clerk 
Maxwell Telescope is operated by The Joint Astronomy Centre on behalf 
of the Particle Physics and Astronomy Research Council of the United 
Kingdom, the Netherlands Organisation for Scientific Research, and 
the National Research Council of Canada.}, mostly between 1993 and 1996. 
The JCMT has a pointing accuracy better than 2$''$ r.m.s. Observations 
of the $\pci$ transition at $\nu$ = 492.161 GHz were made at a resolution 
of 10.2$''$ (HPBW); those of the $J$=2--1 $\13co$ transition with a 
resolution of 22$''$. The observing conditions can be judged from the
total system temperatures (including sky) which are listed in Table~1

All observations were obtained using the Digital Autocorrelator
Spectrograph as a backend. Spectra were binned to various resolutions;
we applied linear baseline corrections only and scaled the spectra to 
main-beam brightness temperatures, $T_{\rm mb}$ = $T_{\rm A}^{*}$/$\eta _{\rm
mb}$. Line parameters were determined by gaussian fitting and by adding
channel intensities over the relevant range. For observations at 492 GHz 
we used $\eta _{\rm mb}$ = 0.43 up to January 1995 and $\eta _{\rm mb}$ = 
0.50 for later observations. For observations at 220 GHz, we used 
$\eta _{\rm mb}$ = 0.69. A list of the observed galaxies and additional 
information is given in Table~1.

\section{Results}

Observational results are summarized in Table~2; a number of 
representative profiles of both [CI] and $J$=2--1 $\13co$ emission is 
shown in Fig.~1. For half of the galaxy sample, the distribution of 
[CI] was mapped beyond the central position (cf. Table 1). The [CI] 
emission maps of NGC~6946 and M~83 have already been published (Israel 
$\&$ Baas 2001); the remaining maps will be discussed in forthcoming 
papers. In the meantime, we have used the information contained in the maps 
to convolve the full-resolution central [CI] intensities (Column~4 of 
Table~2) to [CI] intensities (Column 5) appropriate to the twice larger 
beamsize of the $J$=2--1 $\13co$ observations. We have likewise used
the map information to determine total [CI] luminosities (Column~7) 
integrated over the entire central source extent. 

For all but two of the sample galaxies we have also maps of the $J$=4--3 
$\co$ distribution. Again, we refer to published and forthcoming papers for
a discussion of these observations. For those galaxies that were not
mapped in [CI], we have used the $J$=4--3 $\co$ maps to 
estimate the convolved [CI] intensity and the total luminosity by
assuming identical [CI] and $J$=4--3 $\co$ distributions. This was
shown to be the case for M~83 and NGC~6946 (Petitpas $\&$ Wilson
1998; Israel $\&$ Baas 2001), but we have further verified the 
validity of this assumption for all galaxies that were mapped in both 
[CI] and $J$=4--3 $\co$. Values obtained in this way are given in 
parentheses in Table~2.

In all galaxies mapped, the central neutral carbon peak is well
contained within a radius $R \leq 0.6 $ kpc, often as small as
$R \approx$ 0.3 kpc. For only two galaxies (NGC~5713 and 
NGC~7331) we have no information on extent. In these two cases 
we have listed [CI] luminosities ranging from that observed in
a single beam to that appropriate to the implied maximum source 
diameter of 1 kpc. 

The total [CI] luminosities of the observed galaxies cover a large
range. Quiescent galaxies (NGC~7331, IC~342, Maffei~2, NGC~278, 
NGC 5713) have modest luminosities $\approx 1 \leq L_{\rm [CI]} 
\leq 5 \kkms$ kpc$^{2}$. Galaxies with a starburst nucleus (NGC 253,
NGC~660, M~82, NGC~3628, NGC~6946) have luminosities $ 10 
\leq L_{\rm [CI]} \leq 40 \kkms$ kpc$^{2}$. However, M~83 has only 
$L_{\rm [CI]} = 3.6 \kkms$ kpc$^{2}$, although it is also a starburst
galaxy. The highest luminosities $L_{\rm [CI]} \geq 50 \kkms$ kpc$^{2}$ 
are found in the active galaxies NGC~1068 and NGC~3079.

Interestingly, the ratio of the $\pci$ and $J$2--1 $\13co$ line strengths
exhibits a similar behaviour. The [CI] line is stronger than 
$J$=2--1 $\13co$ in  all galaxies except Maffei~2 and NGC~7331. 
The highest [CI]/$\13co$ ratios of about five belong to the active 
galaxies NGC~1068 and NGC~3079. Generally, the $\pci$ line is weaker than 
the $J$=4--3 $\co$ line, but not by much. In NGC~278, NGC~3079, NGC 4826 
and M~51, the two lines are roughly of equal strength. Only in Maffei~2 
is the [CI] line much weaker.  

\section{Comparison of [CI] and $\13co$ intensities}

%
\begin{figure*}[t]
\unitlength1cm
\begin{minipage}[b]{8.4cm}
\resizebox{7.88cm}{!}{\rotatebox{270}{\includegraphics*{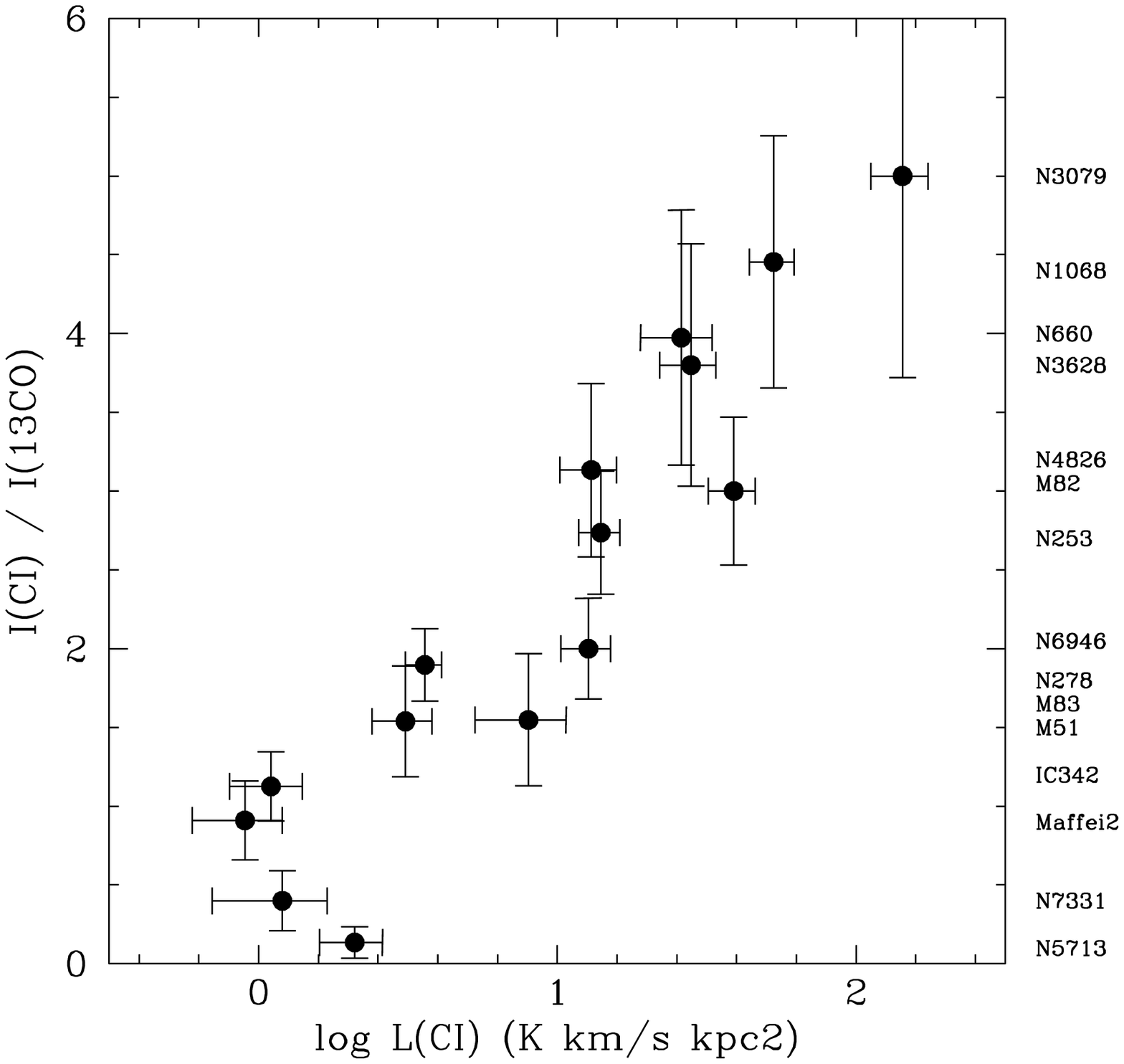}}}
\end{minipage}
\hfill
\begin{minipage}[b]{8.4cm}
\resizebox{8.1cm}{!}{\rotatebox{270}{\includegraphics*{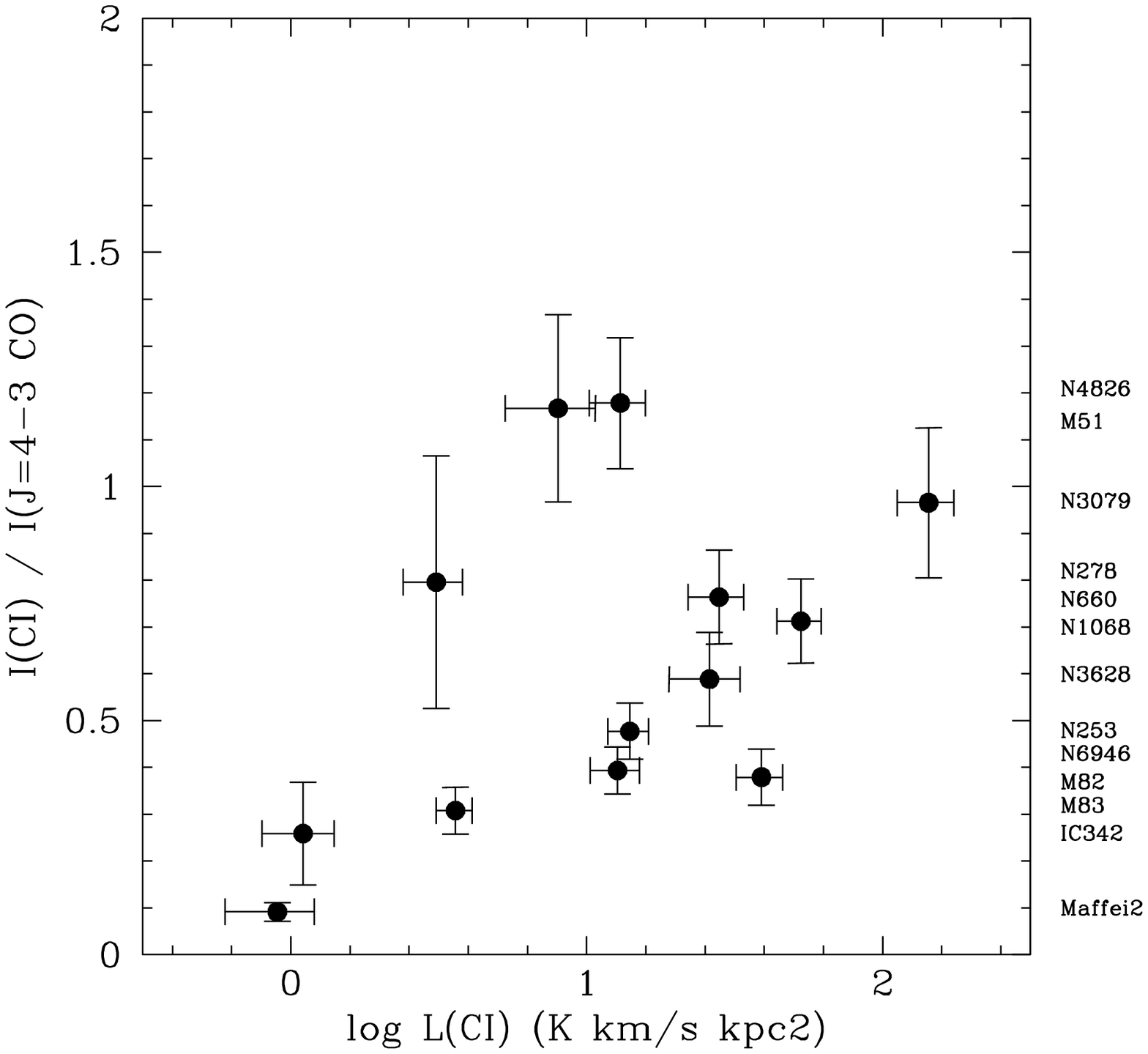}}}
\end{minipage}
\caption[]{Left: [CI]/($J=2-1 \13co$) ratios versus area-integrated [CI] 
luminosity $L[CI]$. Right: [CI]/($J=4-3 \co$) ratios versus $L[CI]$.
The $I$([CI])/$I(\13co)$ ratio appears to be a well-defined function of
log $L$([CI]); $I$([CI])/$I$(4--3$\co$) ratio is not. Galactic sources 
(not shown) would all be crowded together in the lower left corner.
}
\end{figure*}

In various studies of Galactic objects, the similarity of $\pci$ and 
$J$=2--1 $\13co$ intensities and distribution is commented upon. 
Early such studies by the CSO group were reviewed by
Keene et al. (1996). The CSO mapping of the Galactic molecular
cloud complexes M~17 and Oph A show virtually identical line
intensities for [CI] and $\13co$ throughout. This is also found
for most of the Orion Bar and OMC-1. The densest regions of Orion,
however, show increasingly strong $\13co$ emission whereas $\pci$ 
intensities level off, yielding ever lower [CI]/$\13co$ intensity 
ratios down to about 0.4. A similar range of ratios (0.3 - 1.1) was 
found by Jansen et al. (1996) towards the emission/reflection nebula
IC~63. Keene et al. (1996) attributed such low values to the effects 
of enhanced UV radiation in photon-dominated regions (PDR's).
This interpretation finds support in the results obtained
by Plume et al. (1999) and Tatematsu et al. (1999) who used a
reimaging device on the CSO to effectively obtain a larger
beamsize suitable for large-area mapping. Their maps of clouds 
associated with the low-UV sources TMC-1, L~134N and IC~5146 have 
fairly uniform ratios I[CI]/$\13co$ = 1.05 $\pm$ 0.15, as do the 
translucent regions of the dark cloud L~183 observed by Stark et al. 
(1996). In contrast, maps of the molecular clouds associated with 
the high-UV sources W3, NGC 2024, S140 and Cep A yield I[CI]/$\13co$ 
ratios of about 0.5 for the bulk of the clouds. However, even here 
intensity ratios of about unity are found once again at cloud edges. The 
distribution of cloud-edge ratios even has a tail reaching a value of four. 
Only in a few globules associated with the Helix planetary nebula 
(Young et al. 1997) have such relatively high ratios of 3--5 also been found.

Our own data on star formation regions corroborate this: towards
the Galactic HII regions W~58 and ON-1 (unpublished) as well as 
the LMC regions N~159 and N~160 (Bolatto et al. 2000) we find 
intensity ratios [CI]/$J$=2--1 = 0.2--0.6 for the PDR zones associated
with these starforming regions. The two objects
(W-58C and N~159-South) where star formation has not yet progressed to 
a dominating stage, in contrast, yield ratios of about unity. 

As Fig. 2 shows, only a few of the observed galaxy centers
obey the same linear correlations between [CI] and $\13co$
that characterize Galactic clouds. Fully two thirds of the galaxy sample
has $\pci/J=2-1 \13co$ ratios well in excess of unity; the galaxies thus
have much stronger [CI] emission than the $\13co$ intensity and the Galactic 
results would lead us to expect.

The galaxy sample, observed at 15$''$ resolution, discussed by Gerin $\&$ 
Phillips (2000) has only a little overlap with ours, but it shows the same 
effect: more than two thirds of the positions plotted in their Fig.~7
has a ratio [CI]/$\13co$ $\geq 2$. For the galaxy NGC~891, Gerin $\&$
Phillips (2000) observed various positions along the major axis, in addition
to the central region. At the distance of the galaxy, their beam
corresponds to a linear size of 0.5 kpc. Whereas the [CI] intensity 
generally drops with increasing radius, {\it the [CI]/$\13co$ intensity ratio 
increases}, or more specifically, this ratio increases from about 2 at 
the central positions brightest in [CI] to about 4--6 at the disk 
positions weakest in [CI].

Qualitatively, low ratios are expected from regions which have low
neutral carbon abundances. Low neutral carbon abundances will be found
in high-UV environments where neutral carbon will become ionized, and in
environments with high gas densities and column densities. Here, neutral 
carbon disappears because of the concomitant higher CO formation rates
at high densities and the much more efficient CO (self)shielding at
high column densities. Because of its lower abundance, $\13co$ requires
larger column densities for for efficient shielding. Conversely, in 
environments characterized by low gas column densities and mild UV 
radiation fields, such as found in translucent clouds and at cloud 
edges, CO will be mostly dissociated, and most gas-phase carbon may be 
neutral atomic. The resultant relatively high neutral carbon abundance 
will then explain high [CI]/$\13co$ intensity ratios. In this framework, 
our observations and those obtained by Gerin $\&$ Phillips 
(2000) imply that most of the emission from galaxy centers does not come 
from very dense, starforming molecular cloud cores.

\section{[CI]/CO modelling}

%
\begin{figure}[t]
\unitlength1cm
\begin{minipage}[b]{8.4cm}
\resizebox{7.88cm}{!}{\rotatebox{270}{\includegraphics*{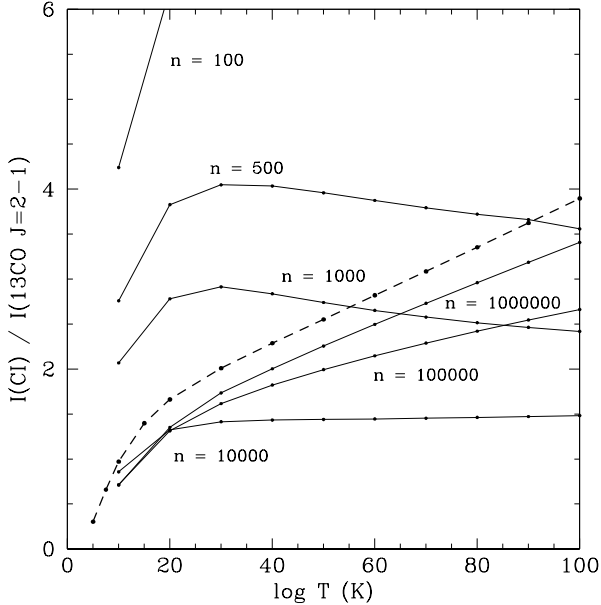}}}
\end{minipage}
\caption[]{Model line ratio $I(CI)/I(J=2-1 \13co$) versus $T_{\rm kin}$ 
resulting for equal column densities $N(CI)$/d$V = N(CO)$/d$V$ 
(solid lines). Curves for a range of total gas densities are marked in 
particles per cc. Dashed line: the ratio $I(CI)/I(J=2-1 \13co$) versus 
$T_{\rm ex}$, likewise requiring equal column densities $N(CI)$/d$V 
= N(CO)$/d$V$ but assuming optically thin LTE conditions. 
}
\end{figure}

In a number of studies (e.g. Schilke et al. 1993, Tauber et al. 1995,
Petitpas $\&$ Wilson 1998) column densities have been calculated assuming 
[CI] and CO emission to occur under optically thin LTE conditions in the
high-temperature limit.
From eqns. (1) and (2) by Tauber et al. (1995), it follows that:

\medskip

$I([CI])/I(\13co) = 0.0063\, A^{12}_{13}\, f(T_{\rm ex})\, N[CI])/N(CO) $

\medskip

\noindent
where 

$f(T_{ex}) = T_{ex}/(e^{7/T_{ex}}+3 e^{-16.6/T_{ex}}+5 e^{-55.5/T_{ex}})$

\medskip

\noindent
while we assume an isotopic abundance $A^{12}_{13} = [\co]/[\13co]$ = 40 
(cf. Mauersberger $\&$ Henkel 1993; Henkel et al. 1998). 
In Fig.~3 we have marked by a dashed line the expected line intensity ratios
for the case $N[CI])/N(CO) = 1$. In order to verify the correctness of the
assumptions of low optical depth and LTE, we have used the Leiden radiative 
transfer models described by Jansen (1995) and Jansen et al. (1994) to 
calculate for gas volume densities ranging from $ 10^{2} \cc$ to $10^{6} \cc$ 
the [CI]/$\13co$ line intensity ratio corresponding to unit column density 
ratios. The calculation was performed for a representative value of the 
column density, $N[CI]$/d$V = N(CO)$/d$V = 1 \times 10^{17}$ cm$^{-2} 
\kms^{-1}$ (cf. Israel $\&$ Baas 2001). Note that in this calculation, 
{\it the temperature parameter is the kinetic temperature $T_{\rm kin}$ 
instead of the excitation temperature $T_{\rm ex}$}. For neutral carbon, 
the two are not very different under the conditions considered, but for 
$\13co$ the excitation temperature is generally much lower than the kinetic 
temperature over most of the relevant range. Under LTE conditions,
the [CI]/$\13co$ ratio continuously increases with temperature $T_{\rm ex}$.
In contrast, the radiative transfer calculation shows that this ratio is 
only weakly dependent on temperature above $T_{\rm kin} \approx 30$ K, and 
in fact decreases slowly with increasing temperature for densities up to
$n(\h2) \approx 10^{4}$. As Fig.~3 illustrates, the assumption of comparable 
excitation temperatures for $\13co$ and [CI] is valid only for very high 
densities $n(\h2) > 10^{6} \cc$ which are unlikely to apply to our observed
sample.

\section{[CI] and CO column densities}

%
\begin{figure*}[t]
\vspace{-3cm}
\unitlength1cm
\begin{minipage}[t]{18.0cm}
\resizebox{18.2cm}{!}{\rotatebox{270}{\includegraphics*{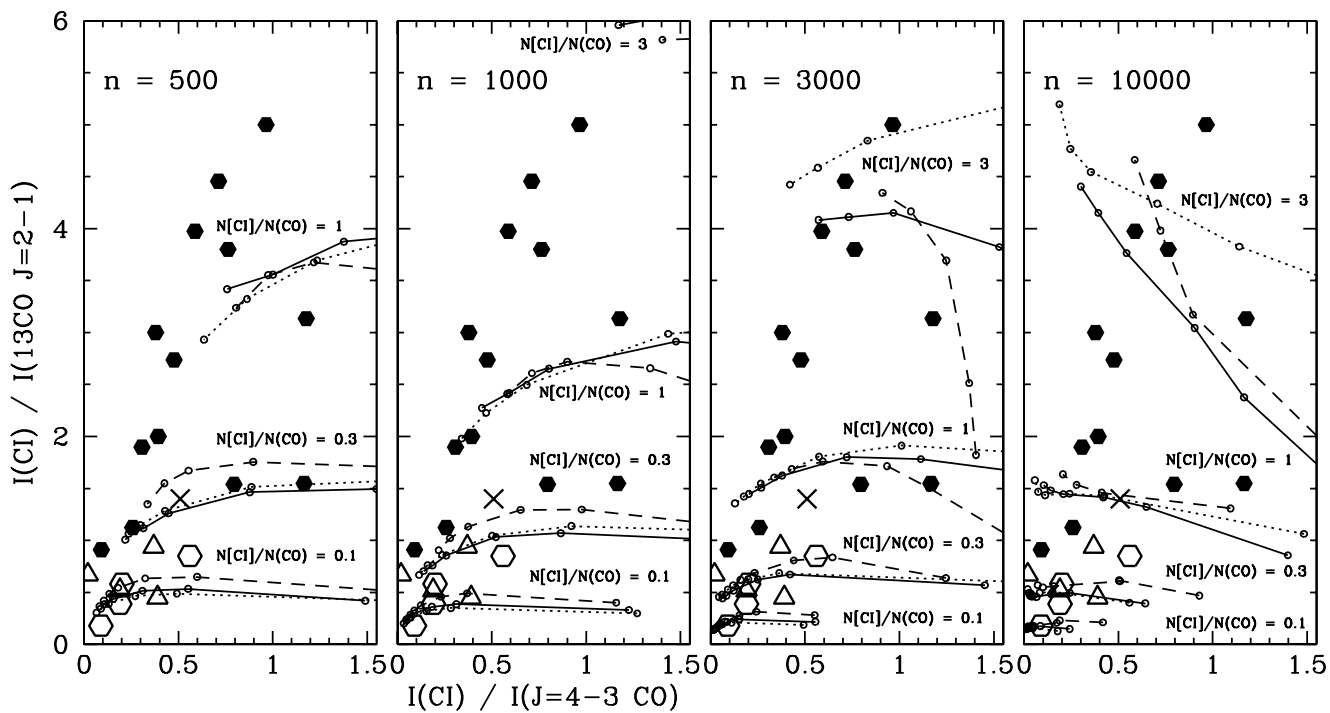}}}
\end{minipage}
\vspace{-1cm}
\caption[]{Observed Line intensity ratios [CI]/$\13co$ versus [CI]/CO 4--3
compared to radiative-transfer model ratios at selected gas densities,
for a range of temperatures and column densities. All model calculations
assume an isotopic ratio $[\co]/[\13co]$ = 40. Galaxy centers are marked by 
filled hexagons, LMC star formation regions N~159 and N~160 by open hexagons, 
Galactic star formation regions W~58 and ON-1 by open triangles and the 
Milky Way Center by a cross. Lines indicate families of column density ratios
N[CI]/N(CO) = 0.1, 0.3, 1 and 3. Within each family, dotted line corresponds to
$N$(CO)/d$V = 3 \times 10^{16} \cm2/\kms$, solid line to $N$(CO)/d$V = 1 
\times 10^{17} \cm2/\kms$ and dashed lines to $N$(CO)/d$V = 3 \times 10^{17} 
\cm2/\kms$. Temperatures of (from left to right 150, 100, 60, 30, 20 and 10 K 
are marked by small open circles on each curve.
}
\end{figure*}

To further investigate the physical conditions characterizing the
central gas clouds that give rise to the observed emission, we have
plotted for our galaxy sample the [CI]/$J=2-1 \13co$ line intensity
ratio versus the [CI]/$J=4-3 \co$ line ratio. For comparison purposes, 
we have added points corresponding to a few Galactic starforming regions 
(White $\&$ Sandell 1995; Israel $\&$ Baas, unpublished), the N159/N160
starforming complex in the Large Magellanic Cloud (Bolatto et al.,
2000), and the Milky Way Center (Fixsen et al. 1999). As the latter do
not list $\13co$ intensities, we have assumed a $J=2-1 \, \co/\13co$
intensity ratio of 8.5, which is the mean value we find for the
galaxies observed by us (Israel $\&$ Baas 1999, 2001, as well as
papers in preparation).

To put the observed points in context, we have used the Leiden radiative
transfer models to calculate the same line intensity ratio in a 
grid with gas densities in the range $n$ = 500 -- 10 000 $\cc$, 
kinetic temperatures in the range $T_{\rm kin}$ = 10 -- 150 K and CO 
column densities $N$(CO)/d$V$ = 0.3, 1.0 and 3.0 $\times 10^{17}$ 
$\cm2/\kms$ respectively. We considered $N$([CI])/$N$(CO) abundance ratios 
of 0.1, 0.3, 1.0 and 3.0 respectively. The results are shown in Fig.~4,
always assuming an isotopic ratio $[\co]/[\13co]$ = 40. Small variations
in the assumed isotopic ratio lead to small shifts in  the various curves
depicted in Fig.~4, mostly along lines of constant temperature.

It is immediately clear from Fig.~4 that the predicted 
[CI]/$\13co$ intensity ratio is roughly proportional to the 
$N$([CI])/$N$(CO) abundance ratio at any given gas-density. Variation 
of the actual CO column density by over an order of magnitude or 
variation of the gas kinetic temperature has very little effect on the 
line intensity ratio except at the highest densities and column densities
where saturation effects caused by high optical depths become dominant.
At given column densities, however, the [CI]/$\13co$ intensity ratio
does depend on the gas-density and is roughly inversely proportional to 
$\sqrt n$. The [CI]/$J=4-3\,\co$ intensity ratio strongly varies as a 
function of gas kinetic temperature and density, as well as column density.

Further inspection of Fig.~4 shows that the starforming regions in
the Milky Way and the LMC are found distributed along curves that
mark neutral carbon versus CO abundances $N$(C$^{\circ}$)/$N$CO)
$\approx$ 0.1--0.3. The galaxy center ratios, in contrast, mostly seem
to imply significantly higher neutral carbon abundances. Only the 
point representing the quiescent bulge of NGC~7331 appears to be
associated with an equally low carbon abundance. Depending on the
assumed value of the total gas density, centers of quiescent galaxies 
are associated with carbon abundances $N$(C$^{\circ}$)/$N$CO) 
$\approx$ 0.3 ($n = 500 \cc$) to 1.0 ($n = 10^{4} \cc$). This is 
consistent with earlier determinations such as $N$(C$^{\circ}$)/$N$CO) 
$\approx$ 0.8 (-0.4, +0.7) for the Milky Way (Serabyn et al. 1994).
In contrast, active galaxies 
have C$^{\circ}$ column densities well exceeding CO column densities
{\it independent of the gas parameters assumed}. The diagonal distribution 
of galaxy points  roughly follows lines of constant kinetic temperature. 
The corresponding temperature value varies as a function of density $n$ 
and column density ($N$): $T_{\rm kin} > 150$ K for $n = 500 \cc$, 
whereas $T_{\rm kin} = 30 - 60$ K for $n = 0.3 - 1.0 \times 10^{4} \cc$, 
$N < 10^{17} \cm2/\kms$. Only the high-density models imply a kinetic 
temperature range covering the fairly narrow dust temperature range
33 K $\leq T_{\rm d} \leq$ 52 K characterizing these galaxy centers
(Smith $\&$ Harvey 1996). 
This can be taken as a suggestion that at least the molecular 
carbon monoxide emission from galaxy centers arises mostly from warm, 
dense gas as opposed to either hot, tenuous gas or cold, very dense gas.
Possible exceptions to this are NGC~278 and in particular NGC~7331, M~51 
and NGC~4826 which occupy positions in the diagrams of Fig.~4 suggesting 
low temperatures  $T_{\rm kin} = 10 - 20$ K and consistent with the
full density range including the highest densities. 

For M~82, Stutzki et al. (1997) estimated from the directly observed
${\,{\rm ^{3}P_{2}-^{3}P_{1}\,[CI]}}/\pci$ line ratio a density $n \geq
10^{4} \cc$ and a temperature $T$ =  50 -- 100 K. This is in very good
agreement with our estimates. However, the $I$([CI])/$I(\13co)$ ratio of 
three suggests an abundance $N$[CI]/$N$(CO) = 2, i.e. four times higher than 
estimated by Stutzki et al. (1997), although not ruled out by their 
results -- see also Schilke et al. (1993).

\section{[CI], [CII] and FIR intensities}

Parameters indicating a warm and dense gas is more or less what is 
expected if the emission arises from photon-dominated region
(PDR's - see e.g. Kaufman et al. 2000). In order to explore this
possibility, we have produced Fig. 5, which presents a comparison of
the $\pci$ and $\pcii$ line and far-infrared continuum intensities; 
this Fig.~5 is directly comparable to Fig.~8 by Gerin $\&$ Phillips 
(2000) who performed a similar comparison. The majority of the
[CII] intensities were taken from 
Stacey et al. (1991), as were the [CII]/FIR ratios. The Milky Way data 
were taken from Fixsen et al. (1999), those for the LMC from Israel et
al. (1996) and Bolatto et al. (2000), and those for NGC~278 from 
Kaufman et al. (2000), who use this galaxy as an example for the 
application of their PDR model calculations. The [CII] measurements 
by Stacey et al. (1991) were obtained in a 55$''$ aperture, i.e. 
much larger than the [CI] beam. We have therefore used the 
area-integrated [CI] intensity, and assumed that all [CII] 
emission from the central source was contained within the 55$''$ 
aperture. As the extent of the central [CI] source is always less than 
this, that seems to be a reasonable assumption. If the [CII] emission 
is more extended than that of [CI], the relevant [CII]/[CI] ratio in Fig.~5 
should be lowered correspondingly. However, we do not believe such a 
correction will change the picture significantly. The [CI]/FIR ratio 
was obtained from the [CII]/[CI] and [CII]/FIR ratio; thus [CI]/FIR may
be somewhat higher than plotted.

%
\begin{figure}
\unitlength1cm
\begin{minipage}[b]{8.4cm}
\hspace{0.35cm}
\resizebox{7.74cm}{!}{\includegraphics*{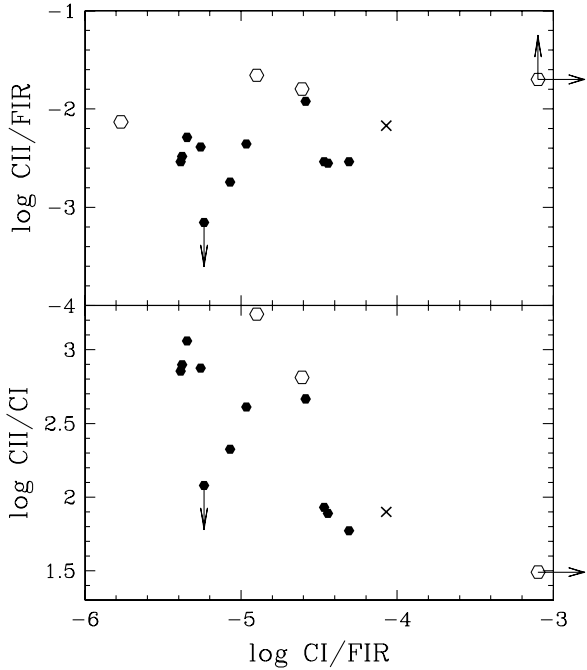}}
\end{minipage}
\caption[]{Ionized carbon [CII] line to far-infrared continuum (FIR) 
ratio (top) and [CII]/[CI] line ratio (bottom) as a function of 
neutral carbon [CI] line to FIR ratio. In both diagrams, the position 
of the Milky Way center is marked by a cross and the positions of 
Magellanic Cloud objects by open hexagons.
}
\end{figure}

In Fig.~5, there is no longer a clear distinction between various types 
of objects such as we found in Fig.~4. Rather, the [CII], [CI] and
FIR intensities define a distribution in which LMC star formation 
regions, low-activity galaxy centers and high-activity galaxy centers 
are all intermingled. Nevertheless, the result shown in Fig.~5 bears
a close resemblance to the the results obtained by Gerin $\&$ 
Phillips (2000). As the [CI]/FIR ratio increases, so does [CII]/FIR, but 
not the [CII]/[CI] ratio which {\it  decreases} with increasing [CI]/FIR.
Qualitatively, this may be explained by PDR process along the line
discussed by Gerin $\&$ Phillips (2000). The horizontal location of the
points in the two diagrams suggest fairly intense PDR radiation fields 
of about 300 to 1000 times the average UV radiation field in the Solar 
Neighbourhood. For the merger galaxy NGC~660 we have only upper
limits (log [CII]/FIR $<$ -3.2, log [CII]/[CI] $<$ 2.1) which place
this galaxy in the same diagram positions as the ultraluminous mergers 
Arp~220 and Mrk~231 observed by Gerin $\&$ Phillips, which correspond 
to strong radiation fields and very high gas densities.

The PDR models shown in Fig.~8 by Gerin $\&$ Phillips provide the highest
[CII]/FIR ratios for model gas densities $n = 10^{3} - 10^{4} \cc$.
Fully half of our observed ratios are well above the corresponding
curves, although they are not quite as high as the ratios observed for 
the three LMC starforming regions. Note that (the limits to) the quiescent 
cloud LMC N159-S in Fig.~5 likewise suggest high densities but only weak 
radiation fields, in good agreement with Israel et al. (1996) and 
Bolatto et al. (2000). For many of the galaxies and for the LMC 
starforming regions, the ratio of [CII] to [CI] intensities appears to be 
higher than predicted by the PDR models considered. For the LMC objects, 
this was already noted and discussed 
by Israel et al. (1996). They explain this situation by an increased 
mean free pathlength of energetic UV photons due the lower metallicity 
of the LMC. However, galaxy centers have, if anything, {\it a higher 
metallicity} (see Zaritsky et al. 1994). A possible explanation for the 
apparently similar behaviour of many galaxy centers may be a greater
degree of filamentary or cirruslike structure. In spite of high metallicities,
this would still allow for an effectively increased penetration depth 
of UV photons. If enhanced exposure results in a significantly larger
fraction of carbon atoms becoming ionized, it would explain higher
[CII] to [CI] emission ratios.   

So far we have assumed homogeneous media, i.e. we have assumed
all CO, [CI], [CII] and FIR emission to originate from the same volume.
This provides in a relatively simple manner good estimates of the
physical paramters characterizing the interstellar medium in the
observed galaxy centers.

The LMC observations, which correspond to linear resolutions one to two
orders of magnitude higher than the galaxy center observations, illustrate
that homogeneity is not the case. The maps shown by Israel et al. (1996) and
Bolatto et al. (2000) show that different locations in the observed 
regions are characterized by strongly different emission ratios
indicating domination by different ISM phases (i.e. neutral atomic, 
ionized, molecular). A similar state of affairs applies to the Galactic
Center region (Dahmen et al. 1998). Ideally, the observations should 
thus be modelled by physical parameters varying as a function of location 
in a complex geometry. Practically, we may approach reality by assuming the 
presence of a limited number of distinct gas components. The analysis of 
multitransition $\co$, $\13co$ and [CI] observations of galaxy centers
such as those of NGC~7331, M~83 and NGC~6946 (Israel $\&$ Baas 1999; 2001) 
suggests that, within the observational errors, good fits to the data can 
be obtained by modelling with only two components: one being dense and 
relatively cool, the other being relatively tenuous and warm. 

The galaxy points in Fig.~5 can all be reproduced by assuming appropriate 
combinations of dense and cold gas (having high [CI]/FIR and [CII]/FIR
ratios) with strongly irradiated gas of lower density (low [CI]/FIR and high
[CII]/[CI] ratios). The distribution of points in Fig.~5 would thus
not directly indicate the physical condition of the radiating gas, but
rather the relative filling factors of the two components.
A similar argument can be made to solve the apparent discrepancy between 
the relatively high kinetic temperatures suggested by Fig.~4 and the 
more modest dust temperatures referred to before. In the same vein, 
a multi-component solution requires somewhat lower beam-averaged
[CI]/CO abundances than suggested by Fig.~4. The dataset presented in 
this paper is, however, not sufficiently detailed to warrant a more 
quantitative analysis such as we have presented for NGC~7331, M~83 and 
NGC~6946 (Israel $\&$ Baas 1999, 2001), and will present for half a dozen 
more in forthcoming papers.

\section{Conclusions}

\begin{enumerate}
\item We have measured the emission from the 492 GHz line corresponding 
to the $\pci$ transition in the centers of fifteen nearby spiral
galaxies. In the same galaxies, we have also measured $J = 4-3 \co$ and 
$J = 2-1 \13co$ intensities for comparison with the $\pci$ line
within the framework of radiative transfer models.

\item Rather unlike Galactic sources, the external galaxy centers have 
[CI] intensities generally exceeding $J$=2--1 $\13co$ intensities, and 
in a number of cases approaching $J$=4--3 $\co$ intensities. 

\item The highest area-integrated (i.e. total central) [CI] luminosities
are found in the active galaxies NGC~1068 and NGC~3079. Slightly
lower luminosities occur in strong starburst nuclei, such as those
of NGC~3628 and NGC~6946. Quiescent and weak-starburst nuclei have
[CI] luminosities an order of magnitude lower.

\item The observed [CI], $\13co$ and $\co$ line ratios, interpreted 
within the context of radiative transfer models, suggest that the bulk 
of the observed emission arises in gas with densities $n \geq 3000 \cc$ 
and kinetic temperatures $T_{\rm kin} \approx 30 - 60$ K. Depending on 
the actual density $n$, most galaxy centers should have abundances 
$N$([CI])/$N$(CO) = 1 -- 3, i.e. [CI] columns just exceeding those of CO. 
Only relatively quiescent galaxy centers such as those of Maffei~2, 
IC~342 and NGC~7331 have abundances $N$([CI])/$N$(CO) $\approx 0.3 - 
1.0$ and are dominated by CO just as the comparison starforming regions 
in the Milky Way and the LMC.

\item The observed [CI] intensities, together with literature [CII] line
and far-infrared continuum data, likewise suggest that a significant
fraction of the emission originates in medium-density gas ($n = 10^{3}-10^{4}
\cc$), subjected to radiation fields of various strengths ranging from
a few times to several thousand times the local Galactic radiation field.

\end{enumerate}

\acknowledgements

We are indebted to Ewine van Dishoeck and David Jansen for providing
us with their their statistical equilibrium calculation models. We also thank 
Maryvonne Gerin and Tom Phillips for communicating to us their neutral
carbon measurements of galaxies well before publication.


\begin{thebibliography}{}
%
\bibitem{} Bolatto A.D., Jackson J.M., Israel F.P., Zhang X., $\&$ Kim S. 
	2000 \apj 545, 234 
\bibitem{} B\"uttgenbach T.H., Keene J., Phillips T.G., $\&$ Walker C.K. 
	1992, \apjl 397, L15
\bibitem{} Dahmen G., H\"uttemeister S., Wilson T.L., $\&$ Mauersberger
	R. 1998 \aua 331, 959
\bibitem{} Fixsen D.J., Bennet C.L., $\&$ Mather J.C. 1999 \apj 526, 207
\bibitem{} Garnett D.R., Skillman E.D., Dufour R.J., et al. 1995 \apj 443, 64
\bibitem{} Gerin M., Phillips T.G. 2000. \apj 537, 644
\bibitem{} Harrison A., Puxley P., Russell A., $\&$ Brand P. 1995 \mnras 
	277, 413
\bibitem{} Henkel C., Chin Y.-N., Mauersberger R., $\&$ Whiteoak J.B. 1998 
	\aua 329, 443
\bibitem{} Hollenbach D.J., $\&$ Tielens A.G.G.M., 1999, Rev. Mod. Phys. 71, 173
\bibitem{} Israel F.P. 1997 \aua 328, 471
\bibitem{} Israel F.P., White G.J., $\&$ Baas F. 1995 \aua 302, 343
\bibitem{} Israel F.P., Maloney P.R., Geis N., et al. 1996, \apj 465, 738
\bibitem{} Israel F.P., Tilanus R.P., Baas F. 1998 \aua 339, 398
\bibitem{} Israel F.P., $\&$ Baas F. 1999 \aua 351, 10
\bibitem{} Israel F.P., $\&$ Baas F. 2001 \aua 371, 433
\bibitem{} Jansen D.J. 1995 Ph.D. Thesis, University of Leiden (NL)
\bibitem{} Jansen D.J., van Dishoeck E.F., $\&$ Black J.H. 1994 \aua, 282, 605
\bibitem{} Jansen D.J., van Dishoeck E.F., Keene J., Boreiko R.T. $\&$ Betz 
	A.L., 1996 \aua 309, 899
\bibitem{} Kaufman M.J., Wolfire M.G., Hollenbach D.J., $\&$ Luhman M.L. 1999
	\apj 527, 795
\bibitem{} Keene J., Blake G.A., Phillips T.G., Huggins P.J., $\&$ Beichman 
	C.A. 1985 \apj 299, 967
\bibitem{} Keene J., Lis D.C., Phillips T.G., $\&$ Schilke P. 1996 in 
	`Molecules in Astrophysics', IAU Symp. 178, ed. E.F. van Dishoeck, 
	p. 129
\bibitem{} Mauersberger R., Henkel C. 1993 Rev. Modern. Astron. 6, 69
\bibitem{} Petitpas, G.R., $\&$ Wilson, C.D. 1998, \apj 503, 219
\bibitem{} Plume R., Jaffe D.T., Tatematsu K., Evans N.J., $\&$ Keene J. 1999
	\apj 512, 768
\bibitem{} Serabyn E., Keene J., Lis D.C., $\&$ Phillips T.G., 1994, \apjl
        424, L95
\bibitem{} Smith B.J., $\&$ Harvey P.M., 1996 \apj 468, 193 
\bibitem{} Sodroski T.J., Odegard N., $\&$ Dwek E., et al. 1995 \apj 452, 262
\bibitem{} Schilke P., Carlstrom J.E., Keene J., $\&$ Phillips T.G. 1993,
	\apjl 417, L67
\bibitem{} Stacey G.J., Geis N., Genzel R., et al. 1991 \apj 373, 423

\bibitem{} Stark R., Wesselius P.R., van Dishoeck E.F., $\&$ Laureijs R.J. 
	1996 \aua 311, 282
\bibitem{} Stutzki J., Graf U.U., Haas S., et al. 1997, \apjl 477, L33
\bibitem{} Tatematsu K., Jaffe D.T., Plume R., Evans N.J., $\&$ Keene J. 1999
	\apj 526, 295
\bibitem{} Tauber J.A., Lis, D.C., Keene J., Schilke P., $\&$ B\"uttgenbach 
	T.H. 1995, \aua 297, 567
\bibitem{} van Dishoeck E.F. 1998, in: {\it The Molecular Astrophysics
	of Stars and Galaxies -- A Volume Honouring Alex Dalgarno}, eds.
	T.W. Hartquist $\&$ D.A. Williams, Oxford University Press, in
	press
\bibitem{} van Dishoeck E.F., $\&$ Black J.H. 1988 \apj 334, 771
\bibitem{} White G.J., Ellison B., Claude S., Dent W.R.F., $\&$ Matheson D.N. 
	1994 \aua 284, L23
\bibitem{} White G.J., $\&$ Sandell, G. 1995, \aua 299, 179
\bibitem{} Young K., Cox P., Huggins P.J., Forveille T., $\&$ Bachiller R. 
	1997 \apj 482, 101
\bibitem{} Zaritsky D., Kennicutt R.C., $\&$ Huchra J.P 1994 \apj 420, 87
%
\end{thebibliography}
\end{document}